\documentclass{PoS}

\title{Energy and System-Size Dependence of Long-Range Multiplicity Correlations from the STAR Experiment}

\ShortTitle{Energy and System-Size Dependence of Long-Range Multiplicity Correlations.}

\author{\speaker{Terence J Tarnowsky (for the STAR Collaboration)}
        \\Purdue University\\
        E-mail: \email{tjt@physics.purdue.edu}}

\abstract{A discussion of results for short and long-range multiplicity correlations (forward-backward) are presented for several systems (Au+Au, Cu+Cu, and pp) and energies (e.g. $\sqrt{s_{NN}}$ = 200, 62.4, and $\approx$ 20 GeV). These correlations are measured with increasing values of a gap in pseudorapidity, from no gap at midrapidity to a separation of 1.6 units ($|\eta|$ = 0.8). For the highest energy, central A+A collisions, the forward-backward correlation strength maintains a constant value across the measurement region. In peripheral collisions, at lower energies, and in pp data, the maximum appears at midrapidity. This result may indicate the possible formation of high density matter for central A+A collisions at $\sqrt{s_{NN}}$ = 200 GeV.}

\FullConference{Critical Point and Onset of Deconfinement
          4th International Workshop\\
		 July 9-13 2007\\
		 GSI Darmstadt,Germany}

\begin{document}

\section{Introduction}

The study of long-range multiplicity correlations has a long history in $e^{+}-e^{-}$ and $p-p(\overline{p})$ experiments \cite{1,2}. More recently, these studies have been expanded to the realm of ultra-relativistic heavy ion collisions at RHIC \cite{3}. Correlations that extend over a large range in pseudorapidity ($\eta$) have the potential to probe the early stages of heavy ion collisions. 
In hadron-hadron experiments, a linear relationship relating the multiplicity in the forward direction ($N_{f}$) to the average multiplicity in the backward direction ($N_{b}$) was found \cite{4}:

\begin{equation}\label{linear} 
<N_{b}(N_{f})> = a + bN_{f} 
\end{equation}

The linear coefficient, \textit{b}, is referred to as the correlation coefficient (or forward-backward correlation strength). The value, \textit{a}, is a measure of uncorrelated particles. For the definition provided in Eq.~(\ref{linear}) the maximum value of the forward-backward correlation strength, \textit{b}, is one. This corresponds to complete correlation between the produced particles. For \textit{b = 0}, all particles are uncorrelated. The relationship from Eq.~(\ref{linear}) can be expressed in terms of the following expectation values \cite{5}:

\begin{equation}\label{dispersion}
b = \frac{<N_{f}N_{b}>-<N_{f}><N_{b}>}{<N_{f}^{2}>-<N_{f}>^{2}} = \frac{D_{bf}^{2}}{D_{ff}^{2}}
\end{equation}
where $D_{bf}^{2}$ and $D_{ff}^{2}$ are the backward-forward and forward-forward dispersions, respectively. This result is exact and model independent \cite{6}. All quantities in Eq.~(\ref{dispersion}) are directly measurable. The large pseudorapidity and angular coverage of the STAR experiment at RHIC provides an ideal means to measure particle correlations with precision \cite{7}.

In phenomenological particle production models, the hadronization of color strings formed by partons in the collision account for the produced soft particles. In the Dual Parton Model (DPM), at low energies, these strings are formed between valence quarks \cite{5}. At higher energies, there are additional strings formed due to the increasing contribution from sea quarks and gluons. In the DPM, this production mechanism is predicted to produce long-range multiplicity correlations in pseudorapidity. A long-range correlation could be an indication of multiple elementary inelastic collisions in the initial stages of an A+A collision \cite{5}. Additionally, the Color Glass Condensate (CGC) model predicts long-range correlations in association with the formation of an intermediate state, the Glasma \cite{8}. The CGC model is a description of high density gluonic matter at small-x. In the CGC model, the incoming Lorentz contracted nuclei are envisioned as two sheets of colliding colored glass \cite{9}. In the initial stage of the collision, the color electric and magnetic fields are oriented transverse to the beam direction (and each other). After the collision there are additional sources formed that produce longitudinal color electric and magnetic fields. This state is referred to as the Glasma, an intermediate phase between the CGC and Quark Gluon Plasma (QGP) \cite{10}. The longitudinal fields provide the origin of the long-range correlation and are similar to the strings as formulated in the DPM.

\section{Data Analysis}

\begin{figure}
\centering
\includegraphics[width=4in]{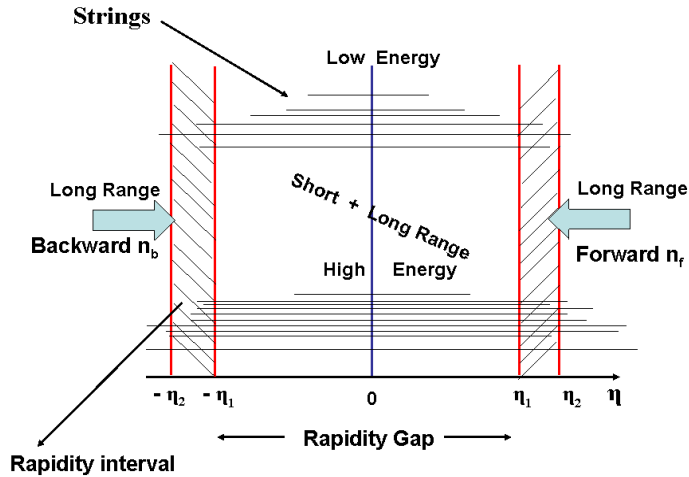}
\caption{\footnotesize{Schematic diagram of the measurement of a forward-backward correlation.}}
\label{fig1}
\end{figure}

The data utilized for this analysis was acquired by the STAR (Solenoidal Tracker at RHIC) experiment at the Relativistic Heavy Ion Collider (RHIC) \cite{7}. RHIC has delivered collisions of the heavy nuclei gold (Au) and copper (Cu), as well as proton-proton collisions. The data discussed here is from Au+Au, Cu+Cu, and pp collisions. For Au+Au collisions, the center-of-mass collision energies ($\sqrt{s_{NN}}$) used are 200 and 62.4 GeV; Cu+Cu, $\sqrt{s_{NN}}$ = 200, 62.4, and 22.4 GeV; and pp, $\sqrt{s_{NN}}$ = 400, 200, and 62.4 GeV. The main tracking detector at STAR is the Time Projection Chamber (TPC) \cite{11}. The TPC is located inside a solenoidal magnet generating a constant, longitudinal magnetic field. For this analysis, data was acquired at the maximum field strength of 0.5 T. All charged particles in the TPC pseudorapidity range $0.0 < |\eta| < 1.0$ and with $p_{T} > 0.15$ GeV/c were considered. This $\eta$ range was subdivided into forward and backward measurement intervals of total width 0.2$\eta$. The forward-backward correlations were measured symmetrically about $\eta = 0$ with varying $\eta$ gaps (measured from the center of each bin), $\Delta\eta =$ 0.2, 0.4, 0.6, 0.8, 1.0, 1.2, 1.4, 1.6, and 1.8 units in pseudorapidity. This is different than the standard two-particle correlation function, which utilizes a relative $\eta$ difference for particle pairs. Figure \ref{fig1} depicts a schematic representation of the forward-backward correlation and its measurement. The collision events were part of the minimum bias dataset. The minimum bias collision centrality was determined by an offline cut on the TPC charged particle multiplicity within the range $|\eta| < 0.5$. The centralities used in this analysis account for 0-10, 10-20, 20-30, 30-40, 40-50, 50-60, 60-70, and 70-80\% of the total hadronic cross section. An additional offline cut on the longitudinal position of the collision vertex ($v_{z}$) restricted it to within $\pm 30$ cm from $z=0$ (center of the TPC). Approximately two million minimum bias events at $\sqrt{s_{NN}}$ = 200 GeV and 0.7 million events at $\sqrt{s_{NN}}$ = 62.4 GeV satisfied these requirements and were used for this analysis. Corrections for detector geometric acceptance and tracking efficiency were carried out using a Monte Carlo event generator and propagating the simulated particles through a GEANT representation of the STAR detector geometry. 

In order to eliminate the effect of statistical impact parameter (centrality) fluctuations on this measurement, each relevant quantity ($\left<N_{f}\right>$, $\left<N_{b}\right>$, $\left<N_{f}\right>^{2}$, and $\left<N_{f}N_{b}\right>$) was obtained on an event-by-event basis as a function of STAR reference multiplicity, $N_{ch}$. A linear fit to $\left<N_{f}\right>$ and $\left<N_{b}\right>$, or a second order polynomial fit to $\left<N_{f}\right>^{2}$ and $\left<N_{f}N_{b}\right>$ was used to extract these quantities as functions of $N_{ch}$. Tracking efficiency and acceptance corrections were applied to each event. Due to statistical limitations, it is not possible to apply corrections for every value of $N_{ch}$. One value for the correction, calculated for each centrality, is applied to the resultant values $\left<N_{f}\right>, \left<N_{b}\right>, \left<N_{f}\right>^{2}$, and $\left<N_{f}N_{b}\right>$ for every event in that centrality. Therefore, all events falling within a particular centrality have the same correction. These were then used to calculate the backward-forward and forward-forward dispersions, $D_{bf}^{2}$ and $D_{ff}^{2}$, binned according to the STAR centrality definitions and normalized by the total number of events in each bin. A strong bias exists if the centrality definition overlaps with the measurement interval in $\eta$ space. Therefore, there are three $\eta$ regions used to determine reference multiplicity: $|\eta| < 0.5$ (for measurements of $\Delta\eta > 1.0$), $0.5 < |\eta| < 1.0$ (for measurements of $\Delta\eta < 1.0$), or $|\eta| < 0.3 + 0.6 < |\eta| < 0.8$ (for measurement of $\Delta\eta = 1.0$). This is possible due to the flat $\eta$ yield of the STAR TPC within the measurement ranges being considered.



\begin{figure}
\centering
\includegraphics[width=4in]{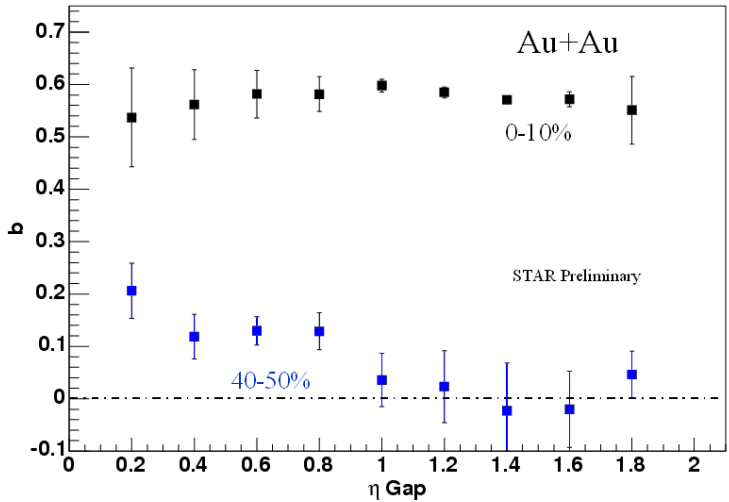}
\caption{\footnotesize{Forward-backward correlation strength, $b$, as a function of the pseudorapidity gap ($\Delta\eta$) in 0-10\% (black squares) and 40-50\% (blue squares) most central $\sqrt{s_{NN}}$ = 200 GeV Au+Au data. If only short-range correlations were present, the expectation is a rapidly falling $b$ value with increasing $\Delta\eta$, as seen in the 40-50\% most central data. The approximately flat behavior of $b$ in 0-10\% data indicates the presence of a long-range correlation.}}
\label{fig2}
\end{figure}

\section{Results}

Figure \ref{fig2} shows the results for the correlation strength $b$ for most central (0-10\%) and mid-peripheral (40-50\%) Au+Au collisions at $\sqrt{s_{NN}}$ = 200 GeV. The short-range correlation (SRC) is defined as the correlation for particle separations less than $\Delta\eta = 1.0$. Long-range correlations (LRC) are considered beyond $\Delta\eta > 1.0$. Figure \ref{fig2} shows that central $\sqrt{s_{NN}}$ = 200 GeV Au+Au data exhibits a strong LRC. The correlation is approximately flat as a function of the pseudorapidity gap. More peripheral (40-50\%) Au+Au data shows only a SRC that decreases rapidly as a function of the pseudorapidity separation. 
It is expected that if only short-range correlations are present (mostly from sources such as cluster formation, jets, or resonance decay), then $b$ would decrease rapidly as a function of $\Delta\eta$. It has been argued that the presence of these long-range correlations at larger than $\Delta\eta = 1.0$ cannot be due to final state effects, and are therefore due to the conditions in the initial state \cite{8}. This may be phenomenologically similar to fluctuations larger than the event horizon being created during an inflationary phase in the early universe that later become smaller than the event horizon \cite{12}.

\begin{figure}
\centering
\includegraphics[width=4in]{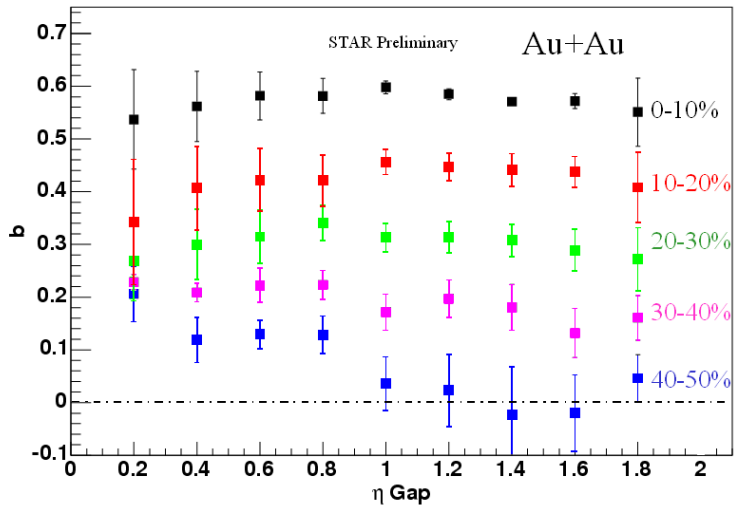}
\caption{\footnotesize{The centrality dependence of the FB correlation strength for Au+Au at $\sqrt{s_{NN}}$ = 200 GeV. Most central (0-10\%) is black; 10-20\%, red; 20-30\%, green; 30-40\%, magenta; and 40-50\%, blue. The long-range correlations persist for the most central Au+Au data, but is similar to the expectation of only short-range correlations at 30-40\% most central data. }}
\label{fig3}
\end{figure}

The evolution of the forward-backward (FB) correlation strength as a function of centrality in $\sqrt{s_{NN}}$ = 200 GeV Au+Au data is presented in Figure \ref{fig3}. There is a smooth evolution in the trend from most central (0-10\%) to mid-peripheral (40-50\%) that shows a decrease of the FB correlation strength as a function of collision centrality. Were only short-range correlations present, the behavior for all centralities should resemble the 40-50\% result. There appear to be long-range correlations for the top 30\% central Au+Au data at $\sqrt{s_{NN}}$ = 200 GeV.

\begin{figure}
\centering
\includegraphics[width=4in]{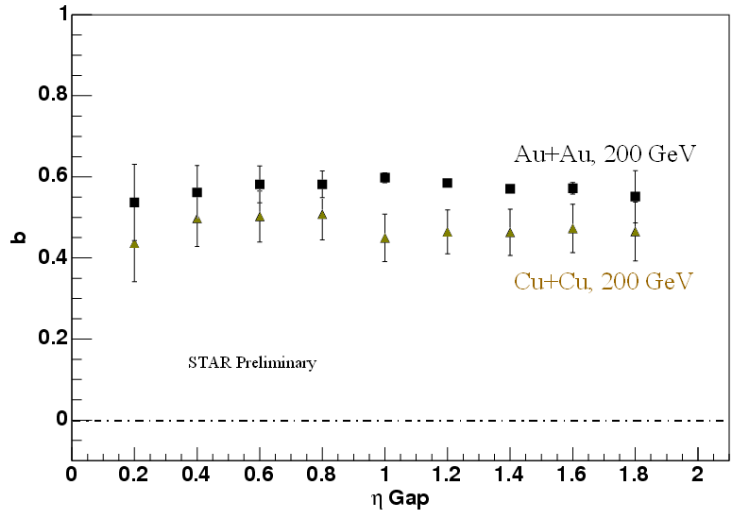}
\caption{\footnotesize{A comparison of the FB correlation strength in 0-10\% most central Au+Au (squares) and Cu+Cu (triangles) data at $\sqrt{s_{NN}}$ = 200 GeV. The long-range correlation seen in Figure \protect \ref{fig2} persists at a slightly lower value in Cu+Cu at the same energy.}}
\label{fig4}
\end{figure}

The system size dependence of the FB correlation strength has been studied in Au+Au, Cu+Cu, and pp collisions. Figure \ref{fig4} shows the comparison of the FB correlation strength for 0-10\% most central $\sqrt{s_{NN}}$ = 200 GeV Au+Au and Cu+Cu. Central $\sqrt{s_{NN}}$ = 200 GeV Cu+Cu exhibits little difference compared to Au+Au. Central Cu+Cu still exhibits a long-range correlation. 

\begin{figure}
\centering
\includegraphics[width=4in]{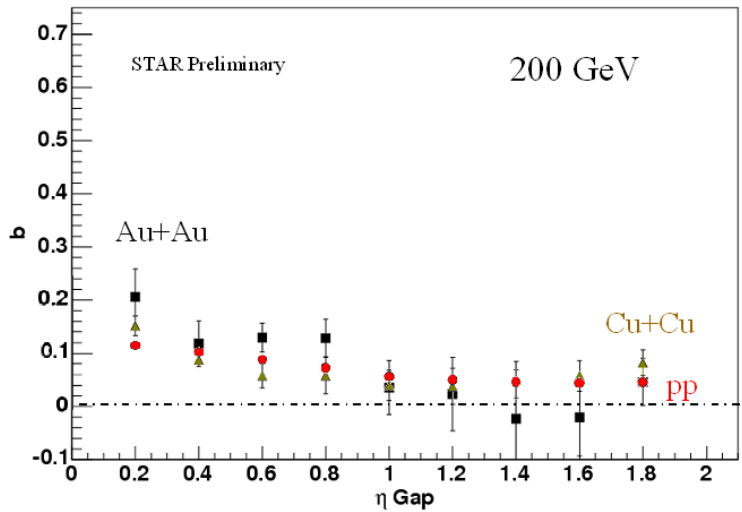}
\caption{\footnotesize{The system size dependence of the FB correlation strength for 40-50\% most central Au+Au (squares) and Cu+Cu (triangles) at $\sqrt{s_{NN}}$ = 200 GeV. Also shown are the results for minimum bias pp (red circles) at the same energy. The mid-peripheral results for all three systems show a similar evolution as a function of the $\eta$ gap.}}
\label{fig5}
\end{figure}

Figure \ref{fig5} shows the system size dependence of the FB correlation strength for mid-peripheral (40-50\%) Au+Au, Cu+Cu, and pp data at $\sqrt{s_{NN}}$ = 200 GeV. As a function of the $\eta$ gap both the heavy ion and pp result exhibit a similar evolution. There is good agreement between all three systems within the calculated errors. The data point closest to mid-rapidity is predominantly driven by short-range correlations. This is exhibited in the ordering of the data near mid-rapidity. The FB correlation strength for Au+Au is largest, followed by Cu+Cu, and pp, following a similar trend in multiplicity.

\begin{figure}
\centering
\includegraphics[width=4in]{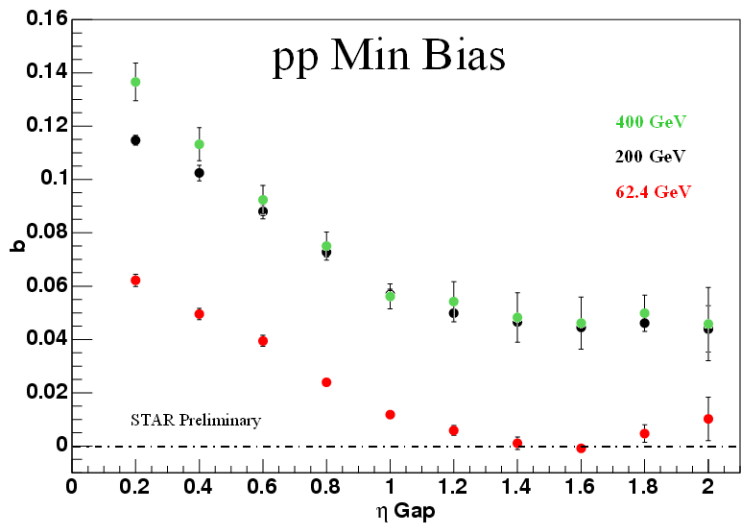}
\caption{\footnotesize{Energy dependence of the FB correlation strength for pp data, as a function of $\Delta\eta$. Three energies are shown: $\sqrt{s_{NN}}$ = 400 (green circles), 200 (black circles), and 62.4 GeV (red circles). The $\sqrt{s_{NN}}$ = 400 and 200 GeV results exhibit close agreement for large values of $\Delta\eta$. The $\sqrt{s_{NN}}$ = 62.4 GeV FB correlation strength is lower than the other two and goes smoothly to $b = 0$.}}
\label{fig6}
\end{figure}

In addition to the system size dependence of the FB correlation strength, the energy dependence has also been studied. Data is available at several different energies for both heavy ions (Au+Au and Cu+Cu) and pp. The energies available for the analysis of the FB correlation strength in pp collisions are $\sqrt{s_{NN}}$ = 400, 200, and 62.4 GeV. For Au+Au and Cu+Cu the available energies are $\sqrt{s_{NN}}$ = 200, 62.4, and $\approx$ 20 GeV. Figure \ref{fig6} is the energy dependence of the FB correlation strength for all three energies of pp. There is good agreement between the FB correlation strength at $\sqrt{s_{NN}}$ = 400 and 200 GeV. The $\sqrt{s_{NN}}$ = 400 GeV data shows a larger short-range correlation, but plateaus at the same value ($b = 0.05$) as the $\sqrt{s_{NN}}$ = 200 GeV result for $\Delta\eta > 1.0$. In contrast, it is seen that at $\sqrt{s_{NN}}$ = 62.4 GeV the FB correlation strength is smaller across the entire $\eta$ region. Unlike the other two energies, the FB correlation strength at this lower energy approaches $b = 0$ at large values of $\Delta\eta$.

The energy dependence of Au+Au and Cu+Cu will be discussed briefly. In the most central data (0-10\%) for both systems at $\sqrt{s_{NN}}$ = 200 and 62.4 GeV, there is a plateau of the FB correlation strength for $\Delta\eta$ > 1.0. However, the plateau at $\sqrt{s_{NN}}$ = 62.4 GeV is $\approx$ 30\% lower in Au+Au and $\approx$ 40\% lower in Cu+Cu. The FB correlation strength for Cu+Cu at $\sqrt{s_{NN}}$ = 22.4 GeV is quantitative agreement w/ the result at  $\sqrt{s_{NN}}$ = 62.4 GeV. The FB correlation for 40-50\% most central data at all energies for both systems is in qualitative agreement. They all exhibit behavior consistent with only short-range correlations.

\section{Discussion}

The Dual Parton Model (DPM) ascribes the long-range correlation to fluctuations in the number of elementary, inelastic collisions, dictated by unitarity \cite{6}. It was shown that the FB correlation strength for pp and mid-peripheral (40-50\%) heavy ion data is dominated by short-range correlations. This contrasts with more central heavy ion data that exhibits a large value of the FB correlation strength for $\Delta\eta > 1.0$. The experimental measurement of a long-range correlation is indicative of multiple partonic interaction in the case of central and semi-central heavy ion interactions.

One possible explanation of the centrality dependence of the FB correlation strength as shown in Figure \ref{fig3} relates to the idea of the formation of a central, partonic core surrounded by a hadronic corona. This has been discussed in terms of hydrodynamic properties and finite formation time of the strongly interacting quark gluon plasma (sQGP) \cite{D1, D2}, as well as attempts to quantify individual contributions from the core and corona \cite{D3}. If the long-range correlation manifests due to partonic effects (as described in the DPM), the partonic core formed in heavy ion collisions would be the source of this correlation. It is expected from geometrical arguments that this partonic core would have a larger volume in central heavy ion collisions, compared to more peripheral overlap. Therefore, the increasing influence of a partonic corona would lead to a larger long-range correlation. The evolution with centrality also agrees with predictions from the DPM and CGC/Glasma phenomenologies. \cite{8}

\section{Summary}

The energy and system size dependence of the FB correlation strength, $b$, has been discussed. It has been studied for three colliding systems: Au+Au, Cu+Cu, and pp. The pp energies range from $\sqrt{s_{NN}}$ = 400 to 62.4 GeV, while those for heavy ions range from $\sqrt{s_{NN}}$ = 200 to $\approx$ 20 GeV. For the most central heavy ion collisions at $\sqrt{s_{NN}}$ = 200 GeV, the value of the FB correlation strength is approximately flat across a wide range in $\Delta\eta$. This behavior slowly disappears as a function of centrality, until $\approx$ 40-50\% most central Au+Au collisions show a FB correlation strength consistent with only short-range correlations. The evolution as a function of $\Delta\eta$ in semi-peripheral data shows a similar behavior across all energies and systems, including pp. The energy dependence of pp shows that for energies greater than $\sqrt{s_{NN}}$ = 200 GeV the FB correlation strength plateaus at a similar (small) value for $\Delta\eta > 1.0$, while $\sqrt{s_{NN}}$ = 62.4 GeV data goes smoothly to $b = 0$. These long-range correlations can be ascribed to multiple elementary inelastic collisions, which are predicted in the Dual Parton Model and the Color Glass Condensate/Glasma phenomenology. The centrality dependence of the FB correlation strength may indicate the increasing dominance of a partonic core in the resulting particle production.

\end{document}